\documentclass{article}
\usepackage{hyperref}
\title{What does a 4.2$\sigma$ discrepancy mean?\\
A brief remark on the statistics of the Muon g-2 Experiment
}
\author{Oliver Passon  \\
University of Wuppertal    \\
School for Mathematics and Natural Sciences\\
passon@uni-wuppertal.de
	}

\date{\today}
\begin{document}

\maketitle

\begin{abstract}
On April 7, 2021 the Muon g-2 Experiment at Fermilab presented its first result which leads to a 4.2$\sigma$ discrepancy between the world average of the muon anomalous magnetic dipole moment and the standard model prediction of this quantity. I take issue with the statistical meaning of this finding.  
\end{abstract}

\section{Introduction}
On April 7, 2021 the Muon g-2 Experiment at Fermilab presented its first results. Combined with the previous  Brookhaven National Laboratory E821 measurement this leads to a new experimental average of the muon anomalous magnetic dipole moment of $a_{\mu}(\mathrm{Exp})=\frac{g_{\mu}-2}{2}=16 592 061(41)\times 10^{-11}$. The difference $a_{\mu}(\mathrm{Exp})-a_{\mu}(\mathrm{SM})=(251\pm59)\cdot 10^{-11}$ has a significance of 4.2$\sigma$ \cite{g-2}. 

Various press releases, news outlets (but also Chris Polly on the special FNAL seminar presenting this result on behalf of the Muon g-2 Collaboration) translate the 4.2$\sigma$ into the claim that the probability of this result being due to chance is just 1 over $40\,000$. This explanation is incorrect.

Given the importance of this beautiful result (which provides indications for BSM physics) I take the liberty to comment on this rather trivial point which is often misrepresented never the less. 

\section{What is the meaning of $p$?}
The above mentioned probability of 1 over $40\,000$ corresponds to the infamous $p$-value in the framework of frequentist  null hypothesis significance testing \cite{glen}.  Famously, on the frequentist conception of probability it is not possible to assign probabilities to hypotheses since they are either true or false (i.e. they are no random variables).\footnote{The frequentist interpretation estimates the probability of an event $x$ by the relative frequency of its occurrence. Hence, there needs to be a repeatable process, $X$, with various possible outcomes.}  To test a hypothesis needs therefor a   detour. The $p$-value is defined as the probability to get the observed (or more extreme) data under the assumption that the so called null hypothesis, $H_0$, is true. In the above case the null hypothesis corresponds to the validity of the standard model.  

There is a huge literature on common misinterpretations and misconceptions surrounding the $p$-value.\footnote{This  debate is particularly heated because the null hypothesis significance test (NHST) is the standard procedure in many fields, including e.g.  psychology or medicine. Here, a $p$-value smaller than 0.05 (which corresponds to $1.96\sigma$ in the Gaussian approximation) is enough to claim a ``statistically significant" result; see Nickerson \cite{nickerson} for a review of this debate and Wasserstein et al. \cite{wasserstein} for recent developments. Given that the NHST is less prominent in physics this debate gets only minor attention here.} Obviously it is incorrect to interpret $p$ as the probability that $H_0$ is true, since $p$ is calculated under the {\em assumption} of $H_0$ being true  (not to mention that the frequentist framework does not allow probability assignments to hypotheses). If we denote the observed data by $D$ (and more extreme data by $D^*$) we may describe the $p$-value symbolically as a conditional probability, $p=P(D^*|H_0)$, which should not be confused with the inverse, i.e. $p\not=P(H_0|D)$. Hence, this misinterpretation has been called the inverse probability fallacy \cite{goodman}.  

However, to claim that the $p$-value corresponds to the probability of getting the observed result by a random fluctuation (as suggested by some in the present case of the g-2 experiment) is incorrect likewise. Note, that the $p$-value has been calculated under the {\em assumption} of a random causation (i.e. assuming $H_0$). Trying to quantify the probability of a random causation needs again to estimate $P(H_0|D)$. That is, we are just dealing with a   variant of the inverse probability fallacy \cite{carver}. 

All this leaves open what the $p$-value {\em does} mean. In order to answer this question it is useful to mention briefly an other common $p$-value misconception, namely the assumption that $p$ is the probability to commit a type I error. This isn't correct either, because a type I error is committed if the (rejected) null hypothesis is true. Hence, the probability of such an error is again related to the probability of a hypothesis being true. However, this misconception is pointing into the right direction still. 

According to  the founding fathers of recent hypothesis testing the meaning of $p$ is the following \cite{np}: If the $p$-value falls below a predefined threshold of, say, $\alpha$ one should {\em act} as if the null hypothesis is false. On this {\em behavioral} strategy it is ensured that {\em in the long run} the type I error rate will be only $\alpha$. One may regret that this meaning implies only little for each individual measurement. However, given that the $p$-value arises in the frequentist framework its meaning has to be frequentist as well. All of the above misconceptions try to derive quantitative information from an {\em individual} trial already, which contradicts the underlying frequentist  (i.e. ``in the long run")  conception.  

As is well known, probability assignments to hypotheses are possible if one moves to a Bayesian framework \cite{yellow}. However, to estimate the probability of the  standard model being valid given the recent Fermilab data needs to include the  prior probability $P(H_0)$ -- which is presumably rather high. 

\section{Conclusion}
On a proper interpretation of the $p$-value obtained by the Muon g-2 Experiment we are justified to {\em act} as if the muon anomalous magnetic dipole moment is deviating from the standard model prediction. However, any quantitative judgment would be premature. As with e.g. the Higgs discovery (which has famously passed even the 5$\sigma$ significance threshold) the trust in a finding is not only a matter of statistical ``significance" but also of an accepted theoretical explanation.\footnote{Note, that the American Statistical Association has recommended to ban the term ``statistical significant" from all scientific publications \cite{wasserstein}.}

\section*{Acknowledgment}
I thank Bryan C. Wills for suggesting this topic.



\begin{thebibliography}{9}
\bibitem{g-2} B. Abi et al. (Muon g-2 Coll.). Measurement of the Positive Muon Anomalous Magnetic Moment to 0.46 ppm. {\em Phys. Rev. Lett.} 2021, {\bf 126}, 141801. 
 \bibitem{glen} G. Cowan. {\em Statistical Data Analysis}. 1998, Oxford: Oxford University Press.
\bibitem{nickerson} R. S. Nickerson. Null hypothesis significance testing: A review of an old and continuing controversy. {\em Psychological Methods} 2000, {\bf 5}, 241--301.
\bibitem{wasserstein} R. L. Wasserstein, A. L. Schirm and N. A. Lazar. Moving to a world beyond ``$p<0.05$''. {\em The American Statistician} 2019, 73(sup1), 1--19.
\bibitem{goodman} S. N. Goodman. A dirty dozen: twelve P-value misconceptions. {\em Seminars in Hematology} 2008, {\bf 45}(3), 135--140.
\bibitem{carver} R. P. Carver. The case against statistical significance testing. {\em Harvard Educational Review} 1978, {\bf 48}(3), 378--399.
\bibitem{np} J. Neyman and E.  Pearson. On the Problem of the Most Efficient Tests of Statistical Hypotheses. {\em Philosophical Transactions of the Royal Society of London. Series A} 1933, {\bf 231}, 289--337.
\bibitem{yellow} G. D'Agostini. {\em Bayesian Reasoning in High-Energy Physics: Principles and Applications}. CERN Report 99-03 (1999).
\end{thebibliography}
\end{document}